# Linear Eckman friction in the mechanism of the cyclone-anticyclone vortex asymmetry and in a new theory of rotating superfluid

S. G. Chefranov

A. M. Obukhov Institute of Atmospheric Physics RAS, Moscow, Russia

schefranov@mail.ru

## Abstract

The observed experimental and natural phenomenon of cyclone-anticyclone vortex asymmetry implies that a relatively more stable and showing a longer life, as well as a relatively more intense mode of rotation with an anticyclonic circulation direction (opposite to the direction of rotation of the medium as a whole) is realized as compared with an oppositely directed rotation of the cyclonic vortex mode. Until now, however, it was not a success to identify a universal triggering mechanism responsible for the formation of the corresponding breaking of chiral vortex symmetry, but, as the laboratory experiments show, such a mechanism should be closely interconnected with the presence of the rotation of the medium as a whole. In this paper we reveal the said linear universal instability mechanism of breaking of chiral symmetry in the sign of vortex circulation in the rotating medium in the presence of linear Eckman friction. Obtained is a condition for the linear dissipative - centrifugal instability (DCI), which leads (only when considering the external linear Eckman friction for an above-threshold value of rotation frequency of the underlying boundary surface of fluid) to the breaking of chiral symmetry in the Lagrangian fluid particle dynamics and the corresponding realization of the cyclone-anticyclone vortex asymmetry. The condition for the DCI is found by considering such a generalization of the Kármán classic solution where the effect of linear external friction is taken into account. A new non-stationary solution to the problem for the disc which carries out weak axial-torsional oscillations in fluid in connection with the experimental data on the rotating superfluid helium-II has been found.

# Introduction

1. **Cyclone-anticyclone asymmetry.** In the atmospheres of the fast-rotating planets (Jupiter, Saturn, Earth), in the ocean and also in the laboratory experiments on rotating fluids observed is the so-called cyclone-anticyclone vortex asymmetry that is manifested itself in explicit predominance of the more intense, stable and long-life vortices just with the anticyclonic direction of circulation which is opposite in the direction of the rotation of medium as a whole [1 – 9]. For example, the Great Red Spot (GRS) of Jupiter has existed for already several hundreds years, and it is just the anticyclonic vortex, one of the first nonlinear models of which is presented in [4]. The similar manifestations of a chiral vortex asymmetry breaking are typical also for the observed anticyclonic vortex lenses in ocean [1, 9] and for the vortices created in the rotating vessels containing fluid in the laboratory experiments [2, 3, 7, 8]. In particular, it is noted in [7, 8], that the chiral-symmetric state of vortex dynamics (when identical in intensity and degree of localization vortices with different rotation directions appear in the system), which exists till the beginning of the vessel rotating, is significantly broken when it rotates in a certain finite frequency range. At the same time, in the rotating vessel observed are an acceleration of the rotation velocity of the anticyclonic vortices and a deceleration of the rotation velocity of the cyclonic vortices, which occupy increasingly more area and show a tendency to their merging [7, 8].

Despite the facts that the vortex chiral asymmetry breaking in the rotating medium is already well-investigated, there is still no consensus on understanding of their common trigger mechanism providing an initiation of such an asymmetry breaking and such a realization of the given cyclone-anticyclone vortex asymmetry [6]. In particular, in [6] noted is the failure in the well-known undertaken attempts to obtain a universal conclusion on the relatively greater instability of the vortices with the cyclonic circulation direction, as compared with the anticyclonic vortices, by deriving from the linear theory of hydrodynamic stability [10 – 12]. On the other hand, none of the previously considered various specific non-linear mechanisms of the realization of the given vortex asymmetry in the rotating medium can be accepted as necessary for the realization of this fundamental phenomenon. For example, for a mechanism related to the nonlinear beta effect (when it is possible to obtain a soliton-like solution for anticyclonic structures only [4]) it has become evident after detection of the anticyclonic circulation type dominance even in f-plane approximation [6, 13]

The offered in [7, 8] nonlinear mechanism of the vortex asymmetry initiation related to the nonlinear Eckman friction effects is seemed to be more general. Besides, as well as for the linear Eckman friction, the presence of the circulation in the vertical plane (i.e., the so-called effects of two-dimensional compressibility [13]) must be taken into account. In particular, it is shown in a theory under development in [7, 8] that as the stationary settled cyclonic component of the vortex field approaches the surface of the fluid, it decays faster and makes a lesser contribution to the observed surface stream than it is the case with the anticyclonic component contribution. But in the theory [7, 8], when considering the limit that corresponds to the linear external Eckman friction, even such a weak vortex asymmetry has already not found (since the very fact of exponential decay upon approaching the fluid surface is realized both for the cyclonic and anticyclonic vortex fields). This weakness of the manifestation of the vortex asymmetry is similar to the observed one in [6] and the linear theory of stability [12] (where both cyclonic and anticyclonic vortex structures are instable, but in the case the latter has lower values of the exponential rates of growth of the corresponding disturbances). In the nonlinear theory [7, 8] there is no such a possibility of determining the conditions leading to the initial non-stationary stage in the formation of the cyclone-anticyclone vortex asymmetry which can be determined only according to the linear theory of the vortex state chiral symmetry stability. At the same time, a consideration of the combination of the fluid rotation with the Eckman friction (both linear and non-linear), that is corresponding to this rotation, has a universal significance for understanding of the mechanism of the cyclone-anticyclone vortex asymmetry phenomenon observed in the rotating medium only. Besides, it is a success in [7] in achieving an agreement of theory based on the nonlinear Eckman friction with the experimental data obtained in the said research for the range of relatively high fluid rotation frequencies only, which correspond to Rossby numbers less than 1.

In the present paper obtained is the hydrodynamic generalization of conclusion [5] made for the simplest linear mechanic system in the form of two-dimensional linear oscillator that has its own oscillation frequency $\omega$ and is investigated in a coordinate system rotating with the frequency $\omega_0$. In paper [5] shown is a possibility of realization of the linear mechanism of the chiral symmetry breaking due to the linear dissipative - centrifugal instability (DCI) of the zero state of the oscillator which may occur only in case of a sufficiently high coordinate system rotation velocity $\omega_0 > \omega$ and only in the presence of a non-zero linear (in oscillator velocity) friction.

In case of the DCI realization, an index of the exponential growth in time for an amplitude of the disturbance reaches its maximum at a certain finite value of the corresponding linear friction coefficient [5]. The chiral symmetry breaking in the DCI is manifested itself in the fact that it is precisely the anticyclonic rotation direction (which is opposite to the direction of coordinate system rotation with frequency $\omega_0$) of the disturbance trajectory that is realized with a sufficiently large time of evolution regardless of the initial conditions. That is, even if the disturbed direction of the movement of the oscillator leaving its zero position of equilibrium is of the cyclonic nature at the initial moment of time (i.e., it coincides with the coordinate system rotation direction), anyway it is just the anticyclonic resultant rotation direction that is formed with time in the limit of large times in case of the DCI realization. The conclusion from paper [5] on the DCI for the linear oscillator is used to interpret the linear mechanism of the observed cyclone-anticyclone asymmetry initiation for description of the Lagrangian fluid particles motion in case of solid-body rotation of fluid with frequency $\omega$ in the coordinate system rotating with frequency $\omega_0$ taking into account the linear Eckman friction that corresponds to linear oscillator linear on velocity friction force.

The application of the DCI mechanism indicated in [5] is also considered in [14] for describing the first stage of tropical cyclone initiation. The present paper shows that such a hydrodynamic consideration [5, 14] is directly linked with the generalization of the Kármán problem solution in case of taking explicitly into account the external linear (Eckman) friction. Determined is the linear mechanism of the DCI chiral symmetrical vortex state that corresponds to this solution and that leads to the observed cyclone-anticyclone vortex asymmetry phenomenon. It is related to the effect of the linear Eckman friction and can be realized only for an above threshold value of the rotation frequency of the fluid vessel. At the same time, the conformity to the experiments data [7] on the vortex asymmetry just within the area of relatively low frequencies (for a Rossby number greater than 1, s. Fig.1) has been obtained that supplements the conclusions in [7] made with regard for the nonlinear Eckman friction effect in the rotating fluid.

**2. Linear (Eckman) friction in rotating superfluid helium-II.**

Nowadays the concept of the L. D. Landau two - fluid model hydrodynamics of superfluid helium-II at a finite temperature is used for interpreting the observed effects of interaction between superfluid helium- II and bodies moving therein (including cases of the proper helium-II rotation in the vessel) [15-17]. The ground for introducing such a concept is the well-known observed properties of helium II showing superfluidity under low temperatures, when, on the one hand, no molecular shear viscosity forces are detected when it flows through narrow slots and capillaries. On the other hand, the measurement of the helium-II viscosity by decaying of the torsional oscillations of a solid disc immersed in the superfluid delivers some finite nonzero values of the surface viscous friction force [15]. The latter refers to any examples of the rotation of helium-II and bodies immersed therein. P.L. Kapitsa [18] detected in his experiments a sharp decrease in the high-efficient heat transfer along the capillary immediately after the completion of the rotation of a cylindrical bar inserted into the capillary that is typical for helium II.

The investigations on the manifestations of the viscous friction forces in the rotating superfluid helium-II show the presence of essential peculiarities in its interaction with the solid disc oscillating therein [16, 17], and it has been considered until now that these peculiarities cannot be manifested themselves in case of interaction of the disc with the normal non superfluid viscous fluid [17, 19]. Mainly this has determined the further formation of the basis of the above macroscopic two -fluid hydrodynamics for the description of the rotating helium II. The conditional nature of such a concept is particularly emphasized in [15], where it is said "it is possible to speak about the superfluid and normal parts of liquid, but it does not mean at all a possibility of a real separation of the liquid into these two parts" and "as with any description of quantum phenomena in the terms of classical theory, it cannot be quite adequate". The most significant inconsistence of the Landau two - fluid superfluid theory arise due to the observations by D. V. Osborne [20] on the free surface of helium- II in the rotating vessel, where just the entire mass of helium- II was found to be involved in the rotating motion as a whole in the contradiction to the two – fluid theory. It should be noted that even before, as noted in [16], such contradictions existed, and they were connected with the necessity of introducing a certain force of friction:"…on the other hand, some experiments seemed to require the existence of yet more frictional forces. And whatever the phenomenological successes of such a friction force, there was no theoretical explanation of its existence" [16].

Richard Feynman offered a solution to this difficult situation related to the contradiction to the two-fluid theory on the zero vortex of superfluid component [21]. His solution is based on the idea about singular vortex filaments that is introduced, however, without taking into account any friction for the rotating helium- II in a vessel. It has been used to apply this two-fluid theory to explanation of the observed effects of interaction between helium- II and the bodies moving therein despite the contradictions to the Kelvin theorem about circulation conservation and to the impossibility of generation of new vortices in an ideal incompressible fluid of constant mass. About this is also noted in paper [17]: "The question on why are the well-known classical Helmholtz-Thomson-Lagrange theorems disapproving the vortex formation in classical ideal liquid not fulfilled is still open-ended". However, avoiding this problem, Feynman gave a quantum-mechanical description of a possibility of the singular vortex lines existence in a superfluid liquid [22]. Indeed, the Feynman theory [22] does not consider the mechanism of formation of the singular vortex filaments in helium- II, but at the same time it does not offer to disapprove Kelvin theorem for quantum liquid. More over in [22] is giving a quantum-mechanical modification of this classic hydrodynamic theory theorem (for the system of the vortex filaments already existing in helium- II) when an invariant quantity of the vortex intensity is supposed to be quantized so that it takes a discrete set of values [16]. The experimental data, on the other hand, explicitly bear witness to the observed evidence of formation of a conventional, but not a singular macroscopic hydrodynamic vortex extending very fast from the surface of the rotating helium II to the vessel bottom [23]. According to the Kelvin theorem, it should indicate a manifestation of a mechanism of friction that takes place between the bottom of the rotating vessel and the helium-II contained therein. It is also noted in paper [16] that at a low helium-II rotation frequency (with periods of more than one minute) the modified two-fluid model theory based on the Feynman singular vortex concept is not consistent with experimental date. This means that it cannot be identified in principal the threshold value of the rotation frequency for the vortex formation and provide a comprehensive understanding of the corresponding mechanism responsible for the macroscopic vortex formation in the rotating superfluid liquid on the basis of the conventional macroscopic two-fluid model hydrodynamics.

Thereupon, in the second paragraph of the present paper it is offered to take into account the effects of the linear Eckman friction for interpretation of the vortex formation mechanism in the rotating helium-II on the basis of the DCI realization, that is considered in [4,14] and the first paragraph herein in connection with the cyclone-anticyclone asymmetry. At the same time, a consideration of the process of the vortex formation in the rotating helium -II, that is an alternative to the two-fluid theory, may be accepted, when it becomes reasonable to take into account even extremely low values of kinematic viscosity coefficient $v$ in helium -II, the order of magnitude of which is $v \cong 10^{-9} см^2/c$ [18]. The corresponding to such value $v$ coefficient of the Eckman external friction force (being linear with respect to fluid velocity) $\alpha = \sqrt{v\omega_0}/h$ (where $\omega_0, h$ are the rotation frequency and the boundary layer thickness, respectively) may be a value of $0.3 \cdot 10^{-4} c^{-1}$ for a certain rotation frequency range [24, 25]. It should be also noted that paper [16] also shows the presence of a relationship between an effect of the linear (by a difference in velocities between the normal and the superfluid component) friction and the helium- II rotation frequency, where the friction is characterized by a coefficient proportional to the vessel rotation frequency $\omega_0$. Besides, it appears that the measured by disc decay effective value $v$ considerably exceeds the value $v$ indicated in [18] and is equal to $v = v_s = (8.5 \pm 1.5)10^{-4} cm^2/\sec$ [16].

In the present paper the generalization of the theory [19] is obtained, where, as opposed to the theory in paper [19], taken into account are the effects of linear Eckman friction and the possibility of a non-coincidence between the fluid rotation frequency $\omega$ far from the disc and the rotation frequency $\omega_0$ of the disc itself. It allows obtaining a new interpretation of the observed peculiarities of the interaction between the rotating helium-II and the solid disc oscillating therein.

## 1. Modification of Kármán problem and DCI

1. Let us consider the equations of hydrodynamics of viscous incompressible fluid to describe a flow above the solid plane disc, rotating with frequency $\omega_0$, with a sufficiently large radius R>> $\sqrt{\omega_0/v}$, when the influence of the disc edge may be considered as low [26]. In the rotating with frequency $\omega_0$ coordinate system these equations are as follows (in the cylindrical coordinate system $z, r, \varphi$) [26]:

$$\frac{\partial V_r}{\partial t} + V_r \frac{\partial V_r}{\partial r} + V_z \frac{\partial V_r}{\partial z} - \frac{(V_\varphi + r\omega_0)^2}{r} = -\frac{1}{\rho_0} \frac{\partial p}{\partial r} + v(\Delta V_r - \frac{V_r}{r^2}) - 2\alpha V_r;$$

$$\frac{\partial V_\varphi}{\partial t} + V_r \frac{\partial V_\varphi}{\partial r} + V_z \frac{\partial V_\varphi}{\partial z} + \frac{V_r V_\varphi}{r} + 2V_r \omega_0 = v(\Delta V_\varphi - \frac{V_\varphi}{r^2}) - 2\alpha V_\varphi;$$

$$\frac{\partial V_z}{\partial t} + V_r \frac{\partial V_z}{\partial r} + V_z \frac{\partial V_z}{\partial z} = -\frac{1}{\rho_0} \frac{\partial p}{\partial z} + v\Delta V_z - g;$$

$$\frac{\partial V_z}{\partial z} + \frac{1}{r} \frac{\partial r V_r}{\partial r} = 0$$

(1)

The system of equations (1) for the velocity and pressure components $V_z, V_r, V_\varphi, p$ is derived assuming the axial symmetry of all considered hydrodynamic fields. This is mean that for all fields in (1) there is no dependence on the angle coordinate $\varphi$ and the corresponding derivatives are equal to zero. In (1) $\rho_0$ is a constant density of fluid, and the z axis is directed perpendicularly to the disc plane and coincides with the direction of the rotation axis and the direction of the gravity force acceleration **g**. The forces of the external (Eckman) friction with the coefficient $\alpha = \sqrt{\nu \omega_0}/h$, which are linear relatively to the horizontal velocity components, are also additionally taken into consideration in the first two equations of the system (1). This coefficient depends on the disc rotation frequency and the typical vertical scale h, which, in particular, may be determined by thickness of the disc. In the classical Karman problem formulation, a stationary solution to the system (1) when $\alpha = 0$ is to be found, and for the radial and tangential velocity field components accepted is the presence of the linear dependence on the radial coordinate (on condition that the vertical component does not depend on this coordinate) [26].

Let us consider the modified Karman problem where the effect of linear external friction with non-zero positive coefficient $\alpha > 0$ is explicitly taken into account. Let the form of solution (1) be the following:

$$V_r = rF(z,t); V_\varphi = rG(z,t); V_z = H(z,t) \qquad (2)$$

From the third equation of system (1), after integration by z, the following equation for determination of the pressure field is derived:

$$p/\rho_0 = -gz + \nu \frac{\partial H}{\partial z} - \frac{H^2}{2} - \frac{\partial}{\partial t}\int dz H(z,t) + \Phi(r,t) \qquad (3)$$

where $\Phi$ is an arbitrary function of integration not depending on z. Below its form will be assumed to be in correspondence with the presence of the fluid uniform rotation with a constant frequency far from the disc, when $\Phi = \frac{\omega^2 r^2}{2}$ as it is the case in [26].

As in [26], let us introduce dimensionless functions and variables using the following relations

$$F = \omega_0 f(\xi, \tau); G = \omega_0 g(\xi, \tau); H = (\nu \omega_0)^{1/2} h(\xi, \tau); \xi = z(\omega_0/\nu)^{1/2}; \tau = t\omega_0 \qquad (4)$$

If substitute (2), (4) into (1) we obtain (taking into account that the relation $f = -\frac{1}{2}\frac{\partial h}{\partial \xi}$ is derived from the last equation of the system (1)) the following system of equations from the first two equations of the system (1):

$$-\frac{1}{2}\frac{\partial^2 h}{\partial \tau \partial \xi} + \frac{1}{4}(\frac{\partial h}{\partial \xi})^2 - \frac{1}{2}h\frac{\partial^2 h}{\partial \xi^2} - (1+g)^2 = -\omega_1^2 - \frac{1}{2}\frac{\partial^3 h}{\partial \xi^3} + \alpha_1 \frac{\partial h}{\partial \xi}, \qquad (5)$$

$$\frac{\partial g}{\partial \tau} - \frac{\partial h}{\partial \xi}(1+g) + h\frac{\partial g}{\partial \xi} = \frac{\partial^2 g}{\partial \xi^2} - 2\alpha_1 g \qquad (6)$$

where $\alpha_1 = \alpha/\omega_0, \omega_1 = \omega/\omega_0$.

The system (6) for the case $\alpha_1 = 0$ exactly corresponds to the system considered in [8] (however, in [8] this system is considered not in the rotating, but in the fixed laboratory coordinate system) in connection with the cyclone-anticyclone vortex asymmetry problem. So, as opposed to [8], in (6) herein the external (Eckman) friction for $\alpha_1 > 0$ has been obviously taken into consideration.

1. Let us find the stationary solution (within the limit of large times, when it is possible to neglect the time derivative terms) to the system (5) and (6) for two unknown functions $h$ and $g$ under the following boundary conditions:

$$g(\xi = 0) = h(\xi = 0) = \frac{dh(\xi = 0)}{d\xi} = 0;$$

$$g(\xi \to \infty) \to g_1 = const; \frac{dh(\xi \to \infty)}{d\xi} \to h_1 = const \tag{7}$$

These boundary conditions on disk surface precisely correspond to those considered by the Karman problem in [26], if taken into account the fact that we are treating it in the coordinate system referred to the rotating disc. As to the second boundary condition at infinity under finite values of the external (Eckman) friction coefficient, the constants in the right parts of the boundary condition (7) at infinity become dependent on the value $\alpha_1$, and the corresponding important difference from [26] is available.

As in [26], let us consider the derivation of an approximate stationary system (6) solution that is linearized by the amplitudes $g, h$ under the boundary conditions (7) if assume that $|\omega_1^2 - 1| \ll 1$. At the same time, (if to take into account the relation $\frac{dh}{d\xi} = -\frac{d^2 g}{d\xi^2} + 2\alpha_1 g$ ), the linearized system (6) comes to a single equation as follows:

$$\frac{d^4 g}{d\xi^4} - 4\alpha_1 \frac{d^2 g}{d\xi^2} + 4g(1 + \alpha_1^2) = 2(\omega_1^2 - 1). \tag{8}$$

The equation (8) should be solved under the boundary condition which follows from (7) taking into account the above linear relations between the functions h and g, when in (7) the constant coefficients are as follows:

$$g_1 = \frac{(\omega_1^2 - 1)}{2(1 + \alpha_1^2)} \approx \frac{|\omega_1| - 1}{1 + \alpha_1^2}; h_1 = \frac{\alpha_1(\omega_1^2 - 1)}{1 + \alpha_1^2} \approx \frac{2\alpha_1(|\omega_1| - 1)}{1 + \alpha_1^2} \tag{9}$$

As a result of solving the equation (8) under the boundary conditions (7), with coefficients from (9), we obtain the following relations for the velocity field components of the fixed flow above the rotating disc in the coordinate system connected with the disc:

$$V_\varphi = r\omega_0 g = \frac{r\omega_0(|\omega_1|-1)}{1+\alpha_1^2}\left[1-(\alpha_1 \sin \xi\lambda_2 + \cos \xi\lambda_2)\exp(-\xi\lambda_1)\right];$$

$$V_r = r\omega_0 f = -\frac{r\omega_0(|\omega_1|-1)}{1+\alpha_1^2}\left[\alpha_1 - (\alpha_1 \cos \xi\lambda_2 - \sin \xi\lambda_2)\exp(-\xi\lambda_1)\right]; (10)$$

$$V_z = (v\omega_0)^{1/2} h = \frac{(v\omega_0)^{1/2}(|\omega_1|-1)}{1+\alpha_1^2}\left[A_0 + 2\xi\alpha_1 + \frac{((\alpha_1\lambda_1-\lambda_2)\cos \xi\lambda_2 - (\alpha_1\lambda_2+\lambda_1)\sin \xi\lambda_2)}{2\alpha_1^2+1}\exp(-\xi\lambda_1)\right]$$

In (10) $\lambda_1 = (\sqrt{\alpha_1^2+1}+\alpha_1)^{1/2}; \lambda_2 = (\sqrt{\alpha_1^2+1}-\alpha_1)^{1/2}$ and

$$A_0 = \frac{1-\alpha_1^2-\alpha_1(2\alpha_1^2+1)}{(2\alpha_1^2+1)(\sqrt{\alpha_1^2+1}+\alpha_1)}.$$

2. When the external Eckman friction coefficient is equal to zero $\alpha_1 = 0$, the solution (10) precisely coincides with the known Karman problem solution that is given in [26]. Under the finite values of this coefficient the solution (10) within the limit $\xi \gg 1$ coincides with the obtained in [14] exact stationary hydrodynamic equation solution (1). This solution corresponds to the solid body rotation with the finite value of helicity (that generalizes the solid body asymptotics of Burgers and Sullivan vortices taking into account the linear external friction) and is considered in [14] in connection with the DCI linear mechanism application for modeling the initial stage of tropical cyclone development. The solution in [14] is obtained for arbitrary values $\omega_1$, when in (2) the similar system (1) solution representation is used under the condition:

$$F = F_0 = const; G = G_0 = const; H = -2F_0 z \qquad (11)$$

It can be shown that Lagrange particles motion (that corresponds to velocity field (2) under the condition (11)) is described in the same way as the two-dimensional oscillator in the rotating coordinate system when friction is linear with respect to velocity (see also [14]). It is just the case that leads to the correspondence between the solution (10) asymptotics and the obtained in [5,14] condition for realization of the DCI of chiral-symmetric state and cyclone-anticyclone vortex asymmetry formation, that, according to (10), is as follows:

$$|\omega_1| < 1 \qquad (12)$$

Actually, from (10) within the limit of large distances from the disc $\xi \gg 1$ we obtain an expression for the velocity field that determines the corresponding ordinary differential equations of Lagrange fluid particles dynamics:

$$V_r = \frac{dr}{dt} = -r\frac{\alpha_1\omega_0(|\omega_1|-1)}{1+\alpha_1^2};$$

$$V_\varphi = r\frac{d\varphi}{dt} = r\frac{\omega_0(|\omega_1|-1)}{1+\alpha_1^2};  \qquad (13)$$

$$V_z = \frac{dz}{dt} = 2z\frac{\alpha_1\omega_0(|\omega_1|-1)}{1+\alpha_1^2}$$

It follows from the first equation (13) that the equilibrium position of a fluid particle, that is localize at r=0, is exponentially unstable only when a finite value of the linear Eckman friction coefficient is positive $\alpha_1 > 0$ and under condition (12), that is at the same time the very condition for the dissipative-centrifugal instability (DCI).

It follows from the second equation in case of the DCI condition realization that the Lagrange particle unstable motion trajectory corresponds to the rotation with a constant negative (when positive disc rotation frequency $\omega_0 > 0$) angular velocity $\frac{d\varphi}{dt} = \frac{\omega_0(|\omega_1|-1)}{1+\alpha_1^2} < 0$, i.e., the trajectory exhibits just the anticyclonic direction. Thus, from the generalization of the Karman problem, taking into account the linear Eckman friction effects, we obtain the confirmation of the conclusion in papers [5, 14] about DCI realization of chiral-symmetrical vortex state and the cyclone-anticyclone asymmetry formation under condition (12).

4. Let us demonstrate that when considering the system (1) solution in the form of (2) with coefficients (11), a DCI condition can be obtained in the same form as (12), but without supposition that $\omega_1 - 1 \ll 1$. If to substitute the (2) and (11) into the first two equations of system (1), two algebraic equations for two unknown coefficients $F_0, G_0$ are obtained:

$$F_0^2 - (G_0 + \omega_0)^2 = -\omega^2 - 2\alpha F_0;$$
$$2F_0(G_0 + \omega_0) = -2\alpha G_0 \qquad (14)$$

From (14) we derive the solution as follows:

$$F_0 = -\frac{\alpha G_0}{G_0 + \omega_0}, \qquad (15)$$

$$x = \frac{G_0}{\omega_0} = -1 \pm \frac{1}{\sqrt{2}}\left[\omega_1^2 - \alpha_1^2 \pm \sqrt{(\omega_1^2 - \alpha_1^2)^2 + 4\alpha_1^2}\right]^{1/2} \qquad (16)$$

The condition for linear (exponential) instability of the state when r=0 is the condition of positivity of the right part of the equation (15), that is observed when x+1>0 (i.e., when considered is only the solution corresponding to the plus sign in front of the square bracket and in front of the square root sign inside the square bracket) and when x<0. The DCI criterion follows immediately from the condition x<0 (12).

For comparing with the paper [7] results, it is possible to use representation (13) for the tangential component of the velocity field (and the corresponding expression for the vertical component of the vortex field $\omega_z = -\dfrac{2\omega_0(1-\omega_1)}{1+\alpha_1^2}$ ), comparing it with the observation evidence shown in Fig. 1 b in [7].

In the paper [7], the laboratory observations of a quasi-two-dimensional vortex flow (that is generated by the system of permanent magnets) were carried out in a rectangular cell that was placed on the rotating platform and filled with a conductive fluid that carried the current of different values. Fig.1 shows the results of the experimental observations which correspond to the different values of the conducted current and are summarized owing to an introduction of a parameter of similarity $\gamma = (I_0/I)^{2/3}$, where $I_0 = 50 mA$ (in Fig. 1 in paper [7], different current values I=30, 40 and 50мА correspond to, respectively, circles, crosses and squares). Besides, in Fig.1, the representation of solution (13) in case of the DCI condition (12) fulfillment corresponds to curve number 2 (on the graph of the vortex field dependence to the third power of the rotation frequency 1/T) , when we obtain the following formula according to (13):

$$\omega_z^3 = -\frac{64\pi^3(1-\omega_1(1/T))^3}{T^3(1+\alpha_1^2(1/T))^3}; \omega_0 = 2\pi/T \text{ при}$$

$\alpha_1 \approx 0.2, \omega_1(1/T = 0.05) \approx 0.3, \omega_1(1/T \geq 0.1) \approx 0.5$.

In Fig.1, the theory of the nonlinear Eckman friction [7] corresponds to two curves under number 1. The lower index 1 is introduced to designate the rotation frequency 1/T and the cube of vortex field $\omega_z^3$ , which is used in [7] to display different current modes investigated in the experiment (for example, $1/T_1 = \gamma(1/T), \gamma = (I/I_0)^{2/3}$, where $I_0 = 50 mA$) in the same figure. The conducted in the present paper comparison with the paper [7] results actually consists of only one case when $\gamma = 1$, when $\langle \omega_1^3 \rangle = \omega_z^3; 1/T_1 = 1/T$ is in Fig.1. It follows from the shown in Fig.1 comparison of the developed theory supported by the experiment and theory [7] that in the range of low rotation frequencies curve 2 from (13) fits well with the experimental data. As a result, the conclusions based on the linear external (Eckman) friction exactly in the relatively low-frequency above threshold range of the DCI realization correlate better with the observation evidence than the results of the nonlinear Eckman friction theory [7].

At the same time, for higher frequencies, as it is shown in Fig.1, theory [7] appears to be adequate enough to the given experimental data. Thus, the present paper demonstrates that under the DCI condition realization (12), the Lagrange particle radial motion, that is determined by the velocity field (2) with coefficients (11), (15), (16), corresponds to the exponential growth of the radial coordinate only under the synchronous rotation of this particle in anticyclonic direction that is opposite to the disc rotation direction. On the contrary, in case of the realization of the cyclonic direction of Lagrange particle rotation (when inequality (12) is broken), the equilibrium position when r=0 for Lagrange particles is already exponentially stable that determines the interconnection between the DCI (12) condition and the mechanism of the observed cyclone-anticyclone vortex asymmetry phenomenon arising in the rotating medium.

## 2. Eckman layer on the rotating fluid and DCI boundary

1. In famous book [26] demonstrated is the coincidence between the obtained approximate Karman problem solution (in the form of (10), but when $\alpha_1 = 0$) and the spiral velocity distribution in the Eckman layer close to the solid boundary above which the rotating fluid flow occurs. It is interesting to derive a generalization of the presented in [26] spiral solution in the Eckman layer (see (4.4. 15), (4.4.16) in [26]) taking into account the hydrodynamic equation terms with the linear external (Eckman) friction, and then compare it with the solution (10) when $\alpha_1 > 0$.

Let us consider the steady-state fluid motion in the boundary layer when the external friction force with the positive coefficient $\alpha$ are additionally taken into account. In this case, we have the following equation for the steady-state values of the velocity field horizontal components inside the boundary layer:

$$-2V_y \omega_0 = \frac{G_x}{\rho_0} - 2\alpha V_x + \nu \frac{d^2 V_x}{dz^2};$$
$$2V_x \omega_0 = \frac{G_y}{\rho_0} - 2\alpha V_y + \nu \frac{d^2 V_y}{dz^2}$$
(17)

where $\omega_0$ is a vertical component of the fluid rotation angular velocity and the steady-state pressure gradient determining the fluid flow is a constant with the components $(-G_x, -G_y)$. These pressure gradient components are expressed by means of the steady-state velocity field horizontal component values $(U_x, U_y)$ in the area above the boundary layer in terms of the following relations:

$$G_x / \rho_0 = 2(\alpha U_x - \omega_0 U_y);$$
$$G_y / \rho_0 = 2(\alpha U_y + \omega_0 U_x)$$
(18)

where the effects of the molecular shear viscosity force are not significant, but the manifestation of the external (linear in relation to velocity) friction force remains important. Relations (17) and (18) when $\alpha = 0, U_y = 0$ precisely coincide with the considered in [26] base equations for describing the fluid that rests relative to the uniformly rotating coordinate axis and is set in motion by a modified pressure uniform horizontal gradient being compensated by the Coriolis force.

The system (17) solution (in case of fulfilling the relations (18)) follows from consideration of zero boundary conditions (on the solid plane z=0 and at infinite large distance from the plane) have the form:

$$V_x = U_x(1 - \cos(\xi\lambda_2)\exp(-\xi\lambda_1)) - U_y \sin(\xi\lambda_2)\exp(-\xi\lambda_1);$$
$$V_y = U_y(1 - \cos(\xi\lambda_2)\exp(-\xi\lambda_1)) + U_x \sin(\xi\lambda_2)\exp(-\xi\lambda_1)$$
(19)

where the values $\lambda_1, \lambda_2$ and the variable $\xi$ precisely coincide with the above determined values of the same quantities which are used in connection with solution (10). When $\alpha = 0, U_y = 0$, relation (19) precisely coincides with the solution obtained in [26] (see (4.4.15), (4.4.16).

In [26] noted is the applicability of this solution for describing the flow near the Earth surface that is accompanied by formation of a twist similar to the Eckman spiral. At the same time, the horizontal pressure gradients may be actually considered as homogeneous at a distance of many kilometers. Their formation may be determined by large-scale cyclones and anticyclones in atmosphere or temperature changes in the horizontal direction due to nonuniform heating of the atmosphere [26].

The characteristic vertical scale that determines the boundary layer thickness (when $\xi \cong \pi$) for the kinematic viscosity molecular coefficient is equal to only 14 m at the terrestrial Poles while its value observed in the atmosphere is much larger (from 500 to 1000 m). It implies a necessity of considering the $\nu$ parameter in (17) as a certain effective kinematic viscosity coefficient under conditions of a small-scale turbulent mixing of the fluid horizontal layers [26]. According to [26], the value of this coefficient may be obtained from the data of observation of the velocity field spiral structure in the boundary layer when comparing it with the corresponding theoretical distribution.

2. Using a finite quantity of the external friction coefficient from (19) we can derive the generalization of the known conclusion on turning wind direction by 45 degrees clockwise with elevation referred to the Earth's surface to the upper boundary of the Eckman boundary layer. It follows from (19) that the relation between the Lagrange fluid particle motion directions near the Earth's surface (when $\xi \to 0$) and at the upper part of the boundary layer is as follows

$$tg\,\varphi_0 = \frac{V_y}{V_x} = \frac{\lambda_1 tg\,\varphi_\infty + \lambda_2}{\lambda_1 - \lambda_2 tg\,\varphi_\infty}; tg\,\varphi_\infty = \frac{U_y}{U_x},$$
(20)

where, as in (10), $\lambda_1 = (\sqrt{\alpha_1^2 + 1} + \alpha_1)^{1/2}; \lambda_2 = (\sqrt{\alpha_1^2 + 1} - \alpha_1)^{1/2}$. Actually, relation (20) generalizes the above noted conclusion on the wind direction turning with elevations. Thus, when $\alpha_1 = \alpha/\omega_0 = 0$ (when in (20) $\lambda_1 = \lambda_2 = 1$) and $U_y = 0$ we obtain the known result $tg\,\varphi_0 = 1$ from (20), when the derived angle $\varphi_0 = 45^0$ does not depend on the rotation frequency $\omega_0$ of fluid as a whole [26]. If to choose (as in [26]) the coordinate axis directions so that the wind direction coincides with the x axis direction, from (20) appears the dependence of the angle $\varphi_0$ on the rotation frequency $\omega_0$, when taking into account the finite value of the linear external (Eckman) friction coefficient $\alpha_1 = \frac{\alpha}{\omega_0}$. Though the limit (on the upper boundary of the boundary layer) wind direction turning angle for $\alpha_1 > 0$ in (20) is smaller than $45^0$, but the direction of the turning just clockwise (i.e. as in an anticyclonic direction of circulation which is also clockwise) remains the same. For example if $U_y = 0; \alpha_1 = 1$ from (20) we have $\varphi_0 = 22.5^0$.

The correspondence (noted in [26]) between solutions (19) and (10) in the considered case of taking into account the linear external friction effects takes place. Indeed, the expressions for the radial and tangential velocity field component in (10) have the same structure as it is the case with solution (19). Besides, the expression for the tangential velocity field component in (10) precisely coincides with the x component of the velocity field in (19) and the y component (but taken with an opposite sign: the necessity to take an opposite sign in a similar correlation is noted in [26], too).

However, the noted coincidence occurs only in case when the following relations hold

$$tg\varphi_\infty = U_y/U_x = \alpha_1; U_x = \frac{r\omega_0(|\omega_1|-1)}{1+\alpha_1^2} \qquad (21)$$

Holding of the first equality in (21) leads to a modification of relation (20). At the same time, in (20) the limit value of the wind turning angle, when elevating from the Earth's surface, depends on only one parameter $\alpha_1 = \alpha/\omega_0$. For example, within the limit $\alpha_1 \ll 1$ it follows from (20), (21) that $\varphi_0 = \pi/4 + \alpha_1/2 + O(\alpha_1^2)$ and taking into account a non-zero external friction leads to increasing in an angle of the mean wind turning with elevation if to compare with the $\varphi_0 = 45^0$ of classical theory [26]. Within the opposite limit $\alpha_1 \gg 1$ we obtain from (20), (21) the following expression $\varphi_0 = \pi/2 - 1/\alpha_1 + O(1/\alpha_1^2)$, and as the external friction coefficient increases, the above angle of wind turning also increases up to the limit value of 90 degrees. When $\alpha_1 = 1$ from (20) and (21) we have $\varphi_0 = 67.5^0$ which is larger than $\varphi_0 = 22.5^0$ (for $U_y = 0$ in (20)) on angle $45^0$.

If $\alpha_1 = 0$, the second of the equalities in (21), that is necessary for establishing the precise correspondence between (10) and (19), also coincides with the equality suggested in [26].

The first equality in (21) leads to an important conclusion that, when taking into account the linear external friction, the y velocity field component outside the boundary layer cannot be guessed independent of the x component and eliminated by an appropriate rotation of the coordinate system. This relation between the velocity field components in (21) may be applied to determine the value of the linear external friction coefficient $\alpha$ on the basis of the angle $\varphi_\infty$ measurement by analogy with the above (and in [26]) method for experimental determination of the kinematic viscosity effective coefficient value.

Actually, subject to the first equality in (21), it follows from (18) that the considered fluid motion is formed exactly by the pressure gradient with the zero component along the x axis (when $G_x = 0$ in (18)). Besides, it follows from (18) that

$$U_x = \frac{G_y}{2\omega_0(1+\alpha_1^2)\rho_0}. \qquad (22)$$

On the other hand, it is also possible to consider the condition $G_x = 0$, when a single-valued relation between the velocity components $U_x$ and $U_y$, coinciding with the first equality in (21), i.e. $U_y = \alpha_1 U_x$, is follow from the hydrodynamic equations in the form of (18). The interrelation between the parameters determining solutions (10) and (19) (when $\omega > 0$) follows from (22) and the second equality in (21) of the form:

$$G_y/\rho_0 = (\omega^2 - \omega_0^2)r \approx 2r\omega_0(\omega - \omega_0). \qquad (23)$$

The dependence on the linear external friction is not present in (23), and the proper expression (23) precisely corresponds to the observed in [26] relation between the modified pressure gradient determining the motion in the Eckman layer and the control parameters in the Karman problem. At the same time, the condition of pressure gradient value negativity in (23) corresponds to the DCI condition $\omega_0 > \omega$, which determine the cyclone-anticyclone vortex asymmetry arising.

3. Let us consider the relation between the DCI condition and the condition of flow energy negativity in the rotating coordinate system in case of a sufficiently high super-critical angular velocity of rotation. Indeed, owing to the well-known representation of the energy in the rotating (round the z axis with the angular velocity $\omega_0$) coordinate system $E = E_0 - M_{0z}\omega_0$ by the energy and the angular momentum in the rest (laboratory) system (denoted by zero index herein), the negativity E<0 is possible under $\omega_0 > \omega_{0th} = E_0/M_{0z}$ [27, 28]. In [28] the above condition of the rotation frequency is considered in connection with the determination of the singular vortex filament formation threshold in the rotating vessel containing helium-II.

Let us note that for the exact solution to the system of hydrodynamic equations (1) in the form of (2), subject to the equalities (11), (15) and (16), the kinetic energy of the respective flow that is treated in the rotating coordinate system is as follows [14]:

$$E = \frac{\rho_0}{2}\left[V_r^2 + V_\varphi^2 + V_z^2 - r^2\omega_0^2\right] = \frac{\rho_0\omega_0^2}{2}\left[r^2 a + 4z^2 b^2\right]; \qquad (24)$$
$$a = b^2 + x^2 - 1; b^2 = \alpha_1^2 x^2/(1+x)^2$$

where the x value is determined in (16). As it is shown in [14], a current expression (24) may be transformed (by making it dependent only on the radial coordinate), using an invariance of value $\Psi = r^2 z = \Psi_0 = const$, that is a stream function for the hydrodynamic equations (1) exact solution in form (2), (11), (15), (16). Besides, it is possible to use the equality $z = \Psi_0/r^2$ in (24) when the expression in square brackets is the function of the radial coordinate only:
$2E/\rho_0\omega_0^2 = r^2 a + 4b^2\Psi_0^2/r^4$. Hence it follows that the necessary condition of the energy

negativity E<0 is to hold an inequality $x^2 < 1$ that provides the DCI realization along with the condition x<0 leading to inequality (12). However, it is evident that this condition is not sufficient for holding E<0 in (24). Such sufficient condition is holding of the inequality a<0 (in case of considering the flow in the plane z=0), which may be expressed as follows:

$$\alpha_1^2 x^2 < (1+x)^3(1-x) \qquad (25)$$

Within the limit $\alpha_1^2 << 1$ it follows from (25) that additionally to the condition x<0 (that leads to the inequality $\omega_1 < 1$ in (12)) we have an additional inequality $x > -1 + (\alpha_1^2/2)^{1/3}$ that leads to the condition of the energy negativity E<0 (which generalizes the DCI condition (12)) of the form:

$$1 > \omega_1 > (\alpha_1^2/2)^{1/3} \qquad (26)$$

In its turn, from (26) it follows that limitations on the rotation frequency $\omega/(\alpha_1^2/2)^{1/3} > \omega_0 > \omega$ both below and above are available. According to experiment [7], the cyclone-anticyclone asymmetry realization is possible not for all rotation frequencies of the vessel containing fluid but have an evident above limitation in this frequency that follows from (26) for the quantity $\omega_0$. On the other hand, the below limitations in the rotation frequency were not specified in [7], as the investigations for the rotation periods larger than 20 sec. were not carried out. Therefore, the experiment [7] evidence does not deny the presence of a lower threshold of a rotation frequency specified by the DCI condition in (12) and (26).

### 3. The oscillation of the disc in helium II and the DCI

1. In the Introduction we have noted the problem related to a difference by many orders in the helium-II viscosity coefficient, when using various methods of its measuring. Indeed, when helium-II flows through narrow slots and capillaries (when the flow velocity in the capillary with a diameter $10^{-5}$ cm may be approximately several centimeters per second) the measured viscosity value does not exceed $10^{-11}$ poise, and when we observe a decay velocity of the torsional axial oscillation of the disc in helium II we obtain the viscosity value variation from $2\times 10^{-5}\ poise$ (close to $\lambda$ point when temperature is 2.19 K ) to $10^{-6}\ poise$ (when temperature is 1 K) [29]. In the present paper we offer to take into account the effect of the external (linear on the velocity) friction. The effect depends on the presence or the absence of the fluid rotation relative to the solid surface. If in case of the absence of the rotation, the coefficient of the external (Stokes) linear friction (this coefficient is proportional to the kinematic viscosity coefficient $\nu$) is $\alpha_0 = \nu/h^2$, and in the presence of the rotation with frequency $\omega_0$ the coefficient of the external (Eckman) friction is $\alpha = \sqrt{\nu\omega_0}/h >> \alpha_0$ (see [24, 25]) when the value $\nu$ is extremely small for helium II.

In paper [19] considered is the issue of low axial and torsional oscillations (with the frequency $\Omega$) of the disc in the viscous incompressible fluid. The disc rotates with this fluid uniformly with the angular velocity $\omega_0$ about the axis coinciding with that axis the disc turns about. This issue is investigated in [19] for interpreting the experimental observation of the same solid disc oscillations in the uniformly rotating helium II in [16,17].

Let us obtain the generalization of the paper [19] conclusions taking into account the linear external (Eckman) friction in the initial hydrodynamic equations (1). At the same time, as in [19], let us consider the system (1) solution linearized representation for low amplitudes of non-stationary disturbances of the velocity and pressure fields caused by low-amplitude disc oscillations in the fluid. As opposed to [19], we use the rotating with frequency $\omega_0$ coordinate system herein, so the disc motion may be expressed as follows: $\varphi = \varphi_0 \exp(i\Omega t)$. We consider the case when it is possible to neglect the influence of the disc end surface on the fluid motion as it is the case with the Karman problem modification in the first paragraph herein. Let the non-stationary solution to the linearized system (1) be similar to (2) when the velocity and pressure fields are as follows:

$$p/\rho_0 = p_0/\rho_0 + \frac{\omega_0^2 r^2}{2} + \frac{r^2(\omega^2 - \omega_0^2)\exp(i\Omega t)}{2} + \nu \frac{dH}{dz}\exp(i\Omega t) \qquad (27)$$

$$V_z = H(z)\exp(i\Omega t); V_r = rF(z)\exp(i\Omega t); V_\varphi = rG(z)\exp(i\Omega t) \qquad (28)$$

By taking into account up to only linear by H, F, G terms under $\omega^2 - \omega_0^2 << \omega_0^2$, we substitute (27), (28) into (1) and obtain the following system after introduction the functions $U_\pm = G \pm iF$:

$$\frac{dU_\pm}{dz} - k_\mp^2 U_\pm = \pm iW; W = \frac{\omega^2 - \omega_0^2}{\nu}; k_\mp^2 = \frac{2\alpha + i(\Omega \mp 2\omega_0)}{\nu} \qquad (29)$$

System (29), when W=0 and $\alpha = 0$, precisely coincides with the considered in [19] system for the fluid motion in the region above the disc. We solve the system (29) under the following boundary conditions:

$$U_\pm(0) = i\varphi_0 \Omega, z = 0; U_\pm(\infty) = U_{\pm\infty} = \mp \frac{iW}{k_\mp^2} = const, \qquad (30)$$

which follows from the condition of non-slip on the disc boundary when we have the equalities F=0 and $G = i\varphi_0\Omega$ in case when z=0. When z=0 the boundary condition coincides with the treated in [19] condition, and the boundary condition at infinity on the z axis shows a difference associated with finiteness of the value W only in the right side of the equations (29).

System (29) solution under the boundary conditions (30) is as follows:

$$U_\pm = \mp \frac{iW}{k_\mp^2} + A_\pm \exp(-zk_\mp); A_\pm = i\varphi_0\Omega \pm \frac{iW}{k_\mp^2};$$

$$k_\mp = \left[\sqrt{\frac{4\alpha^2 + (\Omega \mp 2\omega_0)^2}{4\nu^2}} + \frac{\alpha}{\nu}\right]^{1/2} \pm i\left[\sqrt{\frac{4\alpha^2 + (\Omega \mp 2\omega_0)^2}{4\nu^2}} - \frac{\alpha}{\nu}\right]^{1/2}, \quad (31)$$

where the sign plus before $i$ should appear in case when $x = 2\omega_0/\Omega < 1$, and when x>1 we see the minus sign in the expressions $k_-$ and $k_+$ in (31).

2. Solution (31), as in [19], is used for evaluating the viscous force moment M that influences on the disc and is determined by the value of the vertical gradient of the velocity field tangential component of the disc surface in the following form [26,19]:

$$M = 4\pi\rho_0\nu\int_0^R dr r^2 \left(\frac{\partial V_\varphi}{\partial z}\right)_{z=0} = -2\pi\rho_0\nu\exp(i\Omega t)(A_+ k_- + A_- k_+) \quad (32)$$

The solution (31) representation and the relation $G = \frac{1}{2}(U_+ + U_-)$ were used for deriving the (32) formula. When $\alpha = 0; W = 0$, expression (32) precisely coincides with the representation for M obtained in [19]. Therefore, for finding the dependence of the disc decay decrement $\delta$ and its oscillation frequency $\Omega$ from the fluid rotation frequency $\omega_0$, we use the same representations as in [19] (see [17] also) that are expressed by the imaginary and real part of the value $m = (M/\varphi_0)\exp(-i\Omega t)$ and are as follows:

$$\delta - \delta_0\Omega_0/\Omega = -\pi \operatorname{Im}(m)/I\Omega^2, I = \rho_d\pi R^4 h/2;$$
$$\Omega_0^2 - \Omega^2 = \operatorname{Re}(m)/I \quad (33)$$

where the values $\delta_0$ and $\Omega_0$ correspond to the vacuum values of the decay decrement and the disc oscillation frequency [17]. Besides, in (33) the value $I$ denotes the moment of inertia of the disc of radius R, thickness h and density $\rho_d$. As a result, the following is derived from (32) and (33):

$$\delta - \delta_0\Omega_0/\Omega = \frac{\pi\rho_0}{\rho_d h}\sqrt{\nu/2\Omega}f_\delta(\beta;\varepsilon;x),$$

$$f_\delta = a_- + a_+ + x\varepsilon\left[\frac{a_-\beta\sqrt{x} \pm |1-x|b_-}{\beta^2 x + (1-x)^2} - \frac{a_+\beta\sqrt{x} + (1+x)b_\mp}{\beta^2 x + (1+x)^2}\right],$$

$$a_\pm = (\sqrt{\beta^2 x + (1\pm x)^2} + \beta\sqrt{x})^{1/2}; b_\pm = (\sqrt{\beta^2 x + (1\pm x)^2} - \beta\sqrt{x})^{1/2}, \quad (34)$$

$$x = 2\omega_0/\Omega; \beta = \frac{1}{h}\sqrt{2\nu/\Omega}; \varepsilon_0 = (\omega_2^2 - x^2)/4\varphi_0 \approx \varepsilon x, \omega_2 - x << x,$$

$$\varepsilon = (\omega_2 - x)/2\varphi_0 = x(\omega_1 - 1)/2\varphi_0; \omega_2 = 2\omega/\Omega = \omega_1 x$$

In (34) the representation for the external (Eckman) friction coefficient in the form of $\alpha = \sqrt{\nu \omega_0}/h$ when $2\alpha/\Omega = \beta\sqrt{x}$ is used. In (34) in the numerator of the first term in the square bracket we see the plus sign when x<1 and the minus sign when x>1. In case when $\varepsilon = 0, \beta = 0$, expression (34) precisely coincides with the obtained in [19] formula for the decay decrement.

Similarly, we derive the following representation for the disc oscillation frequency from (32) and the second formula (33):

$$\Omega_0^2 - \Omega^2 = \frac{\pi \rho_0 \Omega \nu}{\rho_d h} \sqrt{\Omega/2\nu} f_\Omega;$$

$$f_\Omega = b_+ \pm b_- + x\varepsilon \left[ \frac{\pm b_- \beta\sqrt{x} - (1-x)a_-}{\beta^2 x + (1-x)^2} + \frac{(1+x)a_+ - b_+ \beta\sqrt{x}}{\beta^2 x + (1+x)^2} \right]$$

(35)

In (35) as well as in (34), the plus sign (before the second term and in the numerator of the first term in bracket) should appear when x<1, and the minus sign when x>1, respectively.

When $\varepsilon = 0$ it follows from (35) that the function $f_\Omega > 0$ is positive for any x and $\beta$. At the finite quantity $\varepsilon$ but this is apparently not the case. For example, when $\beta << 1; \beta\sqrt{x} << |1-x| \leq \beta^\gamma << 1, \gamma < 1$ we obtain the evaluation $f_\Omega \cong \pm \varepsilon/\beta^{(3\gamma/2-1)}$ from (35). Hence it appears a possibility of realization of this function large negative values and the corresponding large value of the disc oscillation frequency under the condition of $2/3 < \gamma < 1$ and also at the DCI condition (12) and $\varepsilon < 0$ when the inequality x<1 is take place. At x>1, the same conclusion is valid at a positive value $\varepsilon > 0$, too, in case when the DCI condition (12) is not fulfilled.

Thus, based on (35), we show a possibility of the realization of a mode where the disc axial-torsional oscillations increase in case when the disc interacts with the conventional rotating fluid owing to the linear (Eckman) friction. It demonstrates that the observed decrease in the oscillation period of the disc (when it have rough surface) in the rotating helium II in [16,17] may be interpreted on the basis of the conventional fluid hydrodynamics even if its viscosity coefficient is as extremely low as $\nu \approx 10^{-9} - 10^{-10} cm^2/\sec$ [18, 29]. Previously, as well in paper [19] (on the basis of which in [17] excluded was the possibility of using the conventional, not two-fluid model, hydrodynamics for interpreting the observed disc and helium-II interaction effects), the possibility of taking into account exactly the external Eckman friction in the helium-II experiments was not considered.

In [16] it is noted that the observed increase in the disc oscillation frequency may be related to the superfluid liquid flow realization in antiphase referred to the disc, which the liquid directly interacts with. Such a concept contradicts essentially the two-fluid superfluidity theory and was criticized in [17]. To interpret the observed disc oscillation frequency increase it is utilized in [17] the representation on the necessity of taking into account elastic properties of the singular vortex filaments existing in the rotating helium-II in case of an above-critical velocity of this rotation ($\omega_0 > \omega_{0\kappa p} \approx 10^{-3} \sec^{-1}$ for the vessel of 1 cm radius [17, 28]).

At the same time, the concept [16] is concordant with the considered above effect related to the external friction mechanism, and, under the DCI condition, leading to the formation of the solid-body anticyclonic fluid rotation being antiphased referred to the disc. The vortex directed oppositely the fluid rotation is also excited under the DCI. At the same time, not only the frequency but also the amplitude of the disc vibration may increase when the decay decrement value in (34) is negative near x=1 under a finite value $\varepsilon$ <0 and a sufficiently small, but not zero value of the parameter $\beta <<1$.

We can also conclude from (34) that, as opposed to [19], the disc oscillation decay decrement value in the conventional fluid may have a local maximum near x=1 under the finite value $\varepsilon$ and sufficiently small, but not zero values of the linear external Eckman friction coefficient determined in (34) with the parameter $\beta$. Fig.2, taken from the paper [19] (see Fig.5 in [17]), shows the observation data (see the upper curve), the paper [19] result (see the lower curve consisting of the two not connected parts), and also added is the curve from (34) when $\varepsilon = 1, \beta = 0.1$ (the middle curve having a local maximum). This figure shows that, as opposed to the paper [19] results, the local maximum observed experimentally in superfluid helium II may appear in the conventional normal fluid, too, but only in case of taking into account the linear external Eckman friction and at a finite value of the parameter $\varepsilon \neq 0$ in (34). The finiteness of the value $\varepsilon \neq 0$ is determined by the possibility to distinguish the angular rotation frequencies $\omega_0$ and $\omega$, for the solid vessel wall and the fluid being far from this solid boundary, respectively.

Such a finite difference between the fluid angular velocities far from the vessel walls may be observed after establishing the equilibrium distribution at sufficiently large times. Indeed, it is noted in the experiment [23] that for approximately 120 sec. the middle part of the fluid remains flat, though during the process of swirling the vessel containing helium II the peripheral parts of the fluid are rapidly entrained by the vessel walls and move up on the wall surface. In the rotating coordinate system, related to such vessel, the indicated behavior of the fluid middle part fully corresponds to the observation of exactly the anticyclonic rotation of the fluid that is similar to the considered in the first paragraph DCI mechanism conditioned by the effect of the external Eckman friction on the vessel bottom.

Indeed, it is noted in [23] that in case of the maximum rotation velocity of 5 revolutions per second possible in this experiment, a conic depression appears in the center of the steady-state meniscus (that is not observed in the conventional normal fluids including helium I [23]) and have been twice transformed into the vortex extending to the vessel bottom. However, these cases of the macroscopic vortex formation have not been recorded on videotape, and the conditions for such a vortex formation remains unexplored. Nevertheless, the obtained in [23] material allows considering the developed in this paper theory as valid instead of the two-fluid theory of helium II which admits the existence of only the microscopic (having an atomic size nucleus) vortex filaments in helium II.

The same conclusion can be made on the basis of Fig.3 where shown are not only the experimental curve, displayed in Fig.2, but also the theoretical curve 4 corresponding to (34) and the curves 2 and 3 corresponding to the theory [17] that proceeds from the singular microscopic vortices conception. As it is noted in the Introduction herein, the conception of these point (in diameter) vortices was introduced by R. Feynman into the Landau two-fluid theory to eliminate the contradictions upon the results of observation of the meniscus (of a depth not differing from that of a conventional normal fluid meniscus) in the rotating helium II [17]. The developed in the present paper theory of the anticyclonic vortex formation owing to the DCI mechanism does not exclude the possibility of such singular vortex filament excitation and does not contradict with the classical theories of ideal fluid hydrodynamics indicated in the Introduction above.

In Fig.3, that was taken from [17] (see Fig.16 in [17]) the curve 4 derived from (34) and given in Fig.2 is shown. However, we had to compress it on the abscissa. Otherwise, the curve 4 maximum would be located outside Fig.3 when $\omega_0 \approx 0.29 \sec^{-1}$. The other given in [17] curves in Fig.3 indicate the following: curve 1 describes the experimental evidence (that correspond to that shown in Fig.2); curve 2 corresponds to the theoretical curve (based on (4.6), (4.21) in [17]) for the rough disc when nonzero value for the kinematic viscosity coefficient of superfluid component is introduced in [17]; curve 3 corresponds to the same formula [17] calculations but for absolutely smooth disc surface under zero superfluid component viscosity.

It follows from Fig.3 that there is no significant qualitative difference between the conclusions of the theory [17] and the present paper theory. Moreover, the absence of the curve 3 local maximum in case of the disc ideal smooth surface and its presence for curve 2 in case of a finite viscosity exactly of the superfluid component indicate the qualitative conformity to the developed herein conception of the linear external friction against the boundary solid surface that is valid both for the conventional fluid and the helium II.

Thus, the conclusions derived from (34) for the conventional fluid, taking into account the linear external Eckman friction effects, may be a quite sufficient alternative to the two-fluid theory in interpretation of the effects of interaction between a superfluid liquid (helium II) and a solid body moving relative to the fluid.

## Conclusions

Taking into account the extremely low kinematic viscosity coefficient values in helium II, the modes with very large Reynolds numbers [18] may be realized even in flow through a thin capillary. Such Reynolds numbers are typical for the flows in geophysical hydrodynamics in the region of planetary boundary layer, i.e., the Eckman layer. Therefore the application of the linear external Eckman friction to the description of the solid disc and helium II interaction seems to be as justified as it is the case with geophysical hydrodynamics [24, 25]. The obtained results indicate the importance of taking into account the linear external friction effects for all physical problems, where treated is the hydrodynamic or gas-dynamic motion of a continuous medium just in case of the presence of any solid or elastic (as in the case of blood dynamic in cardio – vascular system) boundaries in this medium.

At the same time, when traditionally introducing viscous effects into hydrodynamics consideration, this is not taken into account, since usually investigated are only volumetric, but not boundary dissipative factors, that are determined by the velocity field space variability, but not by the medium velocity itself at a given point of space [15].

On the other hand, I. Newton [30] empirically introduced exactly the linear in relation to the body velocity friction force (known as Stoke's force [15]) for determination of a force acting on a body moving in fluid, that transforms into the quadratic in relation to the velocity force in case of turbulent flow around a body. However, for determination of the velocity field structure of the fluid flowing around the solid boundary layer no extra force proportional to the fluid velocity and taking into account such solid boundary presence, is introduced into the hydrodynamic equations. Usually considered are only the effects of viscous interaction between different fluid layers determined by the velocity gradient values and the boundary conditions of adhesion to the solid surface confining the fluid.

The present paper demonstrates the necessity for a modification of this traditional approach. Owing to considerations of the linear external friction in the investigation of the processes of interaction between hydrodynamic flows and solid boundary, both the experimentally observed vortex asymmetry (curve 2 in Fig.1) and the new interpretation of interaction between the rotating superfluid helium II and solid boundary (Fig.2, 3) may be explained. In this case, the traditionally used formal concept of the two-fluid hydrodynamics in helium-II is no longer necessary.

Thereupon, it is interesting to obtain the similar generalization of the known exactly solvable problems of hydrodynamics like determination of the Stoke's force acting on a solid sphere slowly moving in fluid, Hagen-Poiseuille flow in a round tube, etc. where it seems reasonable to take into account the effects of the linear in relation to velocity external friction as it is done in the present paper for the Karman and Eckman problem and for the description of the interaction between helium II and the solid disc oscillating therein. It may give new view on the many fundamental and apply problems of hydrodynamic (for example to the problem of high energy effectiveness of cardio – vascular system job).

It should be separately stressed that the obtained results may be used for solving issues of identification of the cyclonic and anticyclonic vortex activity role in the formation of global climatic disturbances [31, 32]. It is also important to compare the distribution (19) based on the external friction consideration with the conclusion on the Eckman boundary layer spiral structure made in [33], based on the spiral theory of turbulence. Herewith, the distribution derived in [33] may precisely coincide with (19) and give the same distribution of the mean wind with elevation if the relation $\alpha = k_h \omega_0 / k; \nu = (k^2 + k_h^2)/k$ takes place between the friction coefficients introduced above and the semiempirical coefficients of turbulent viscosity $k, k_h$ introduced in [33]. Such relations provide for better understanding of the physical meaning of the turbulent viscosity coefficient.

The conclusions obtained in the present paper are based on the development of the DCI theory offered in [5], where in addition to establishing the proper fact of the DCI existence (that is known for a long time as secular instability specified by the friction forces [34]) for a linear two-dimensional oscillator in the rotating coordinate system, established is also the new fact that the DCI realization is linked with the chiral symmetry breaking. It seems to escape notice in [34] and previous publications.

Let us also note the similarity between the introduction of the linear in relation to velocity Eckman or Newton friction and consideration both of the deterministic [35] and the random non-stationary [36] factors in the hydrodynamic equations when the corresponding linear friction coefficient has the same dimension as some effective frequency of the system.

This work was the supported by the Program of Presidium of Russian Academy of Science "The Fundamental Sciences to Medicin"2014.

I would also like to express many thanks to I.I. Mokhov and O.G. Chkhetiani for useful discussions as well as to F.A. Pogarsky and A.G. Chefranov for their assistance in figures preparation.

Fig. 1

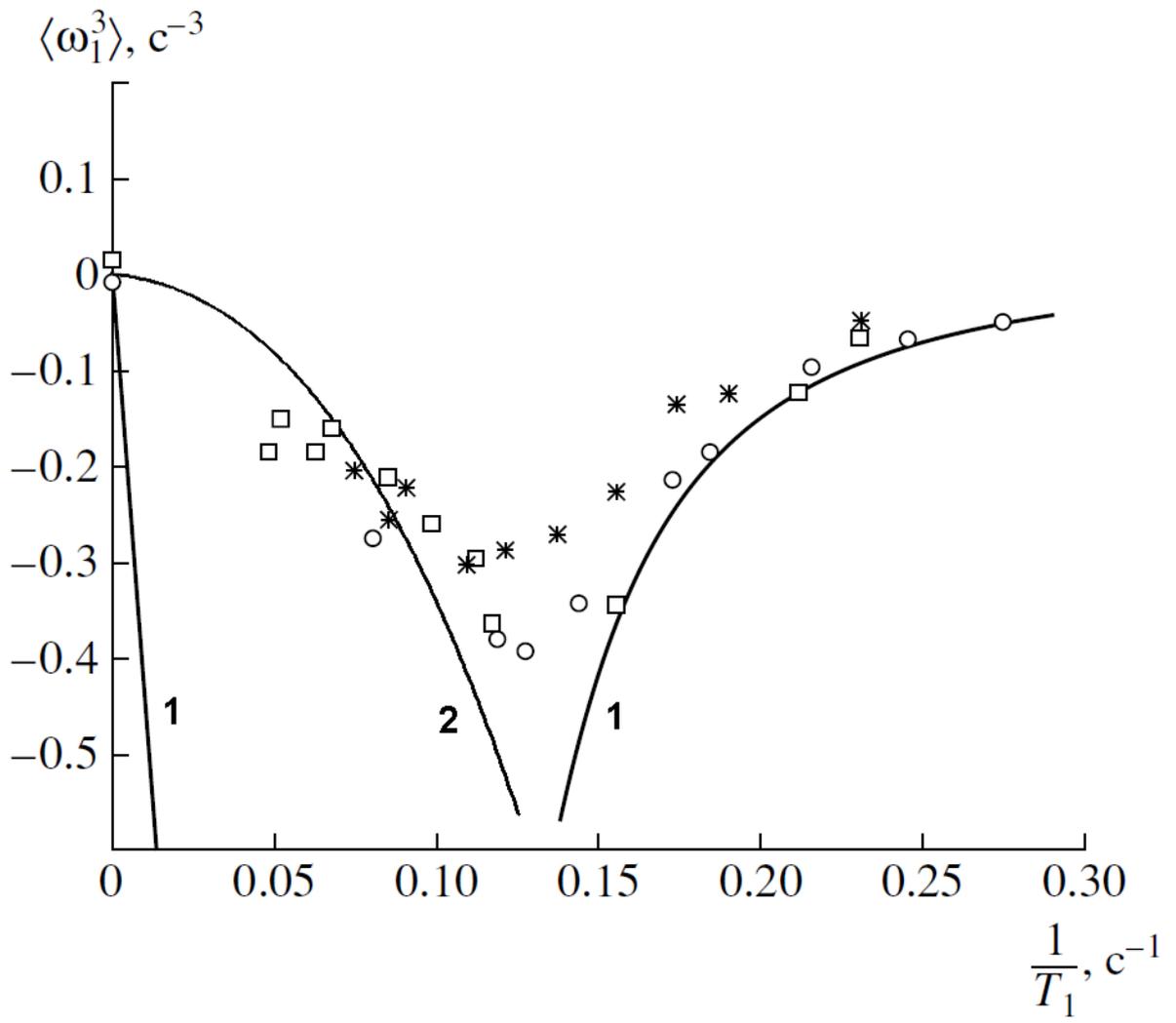

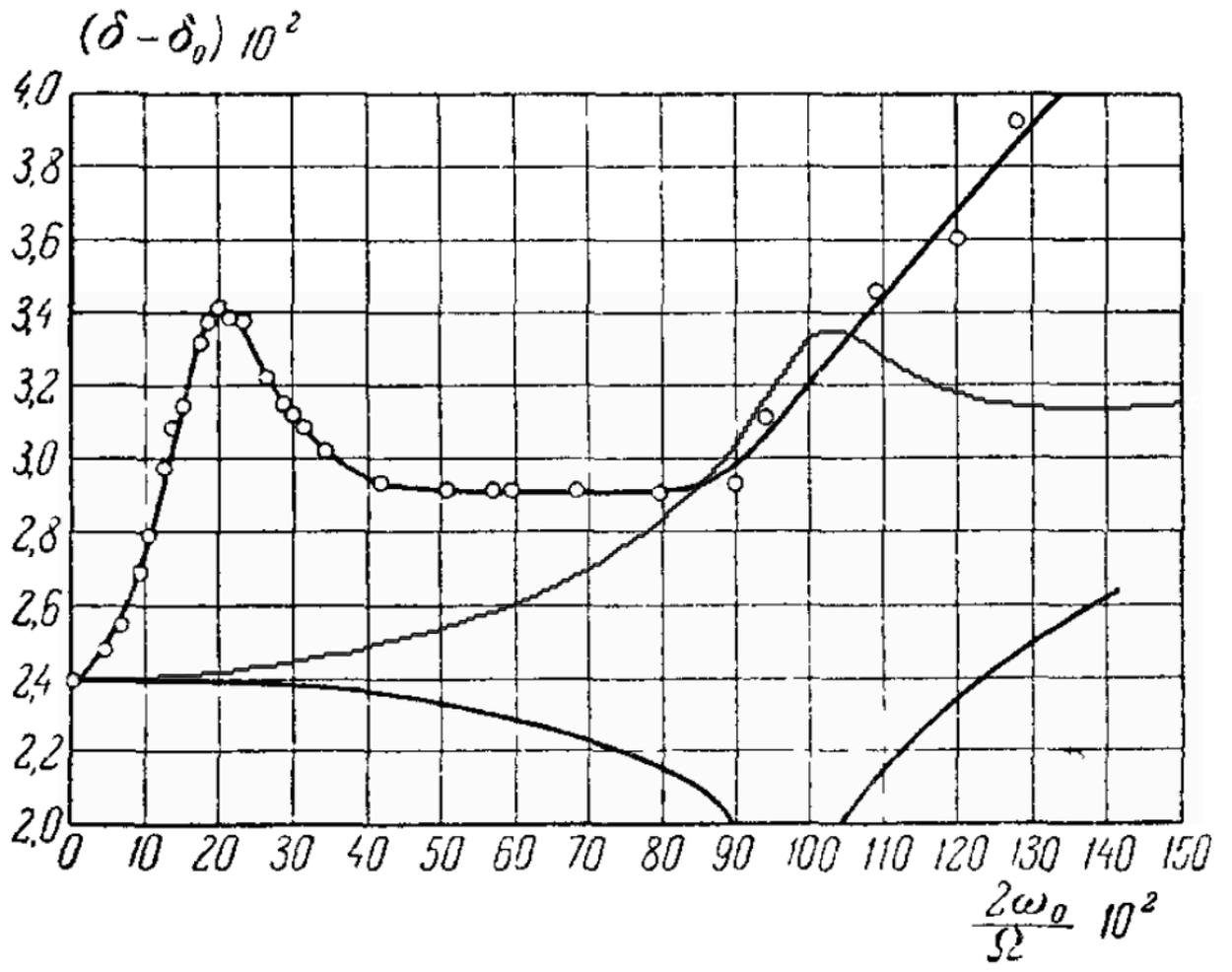

Fig.3

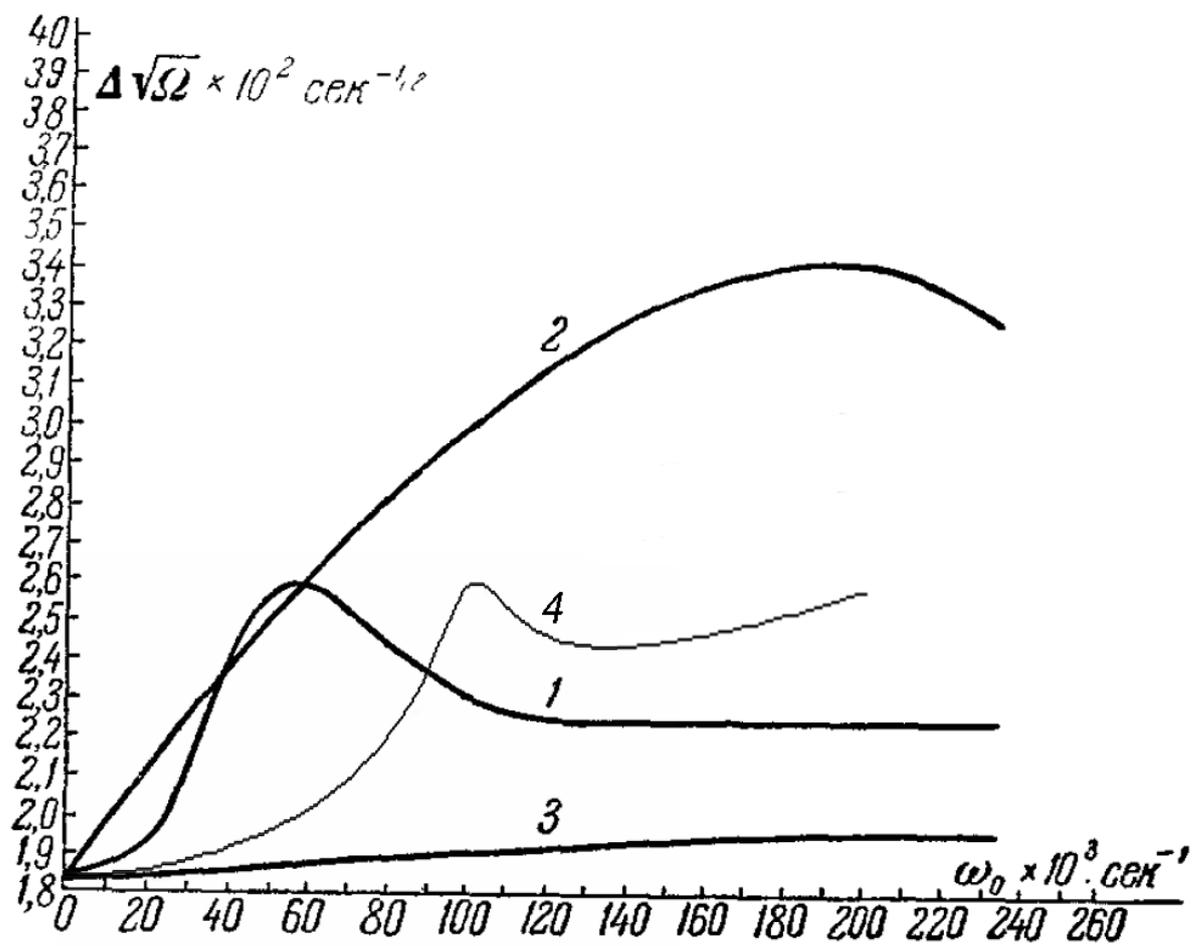

# Литература